\documentclass[a4paper,11pt]{article}
\usepackage{jcappub}

\usepackage{graphicx}
\usepackage{dcolumn}
\usepackage{bm}
\usepackage[nointegrals]{wasysym}
\usepackage{subfigure} 
\usepackage{float}
\usepackage{lineno}

\usepackage{amsmath}
\usepackage{amssymb}

\usepackage[T2A]{fontenc}
\usepackage[utf8]{inputenc}
\usepackage[english]{babel}

\DeclareMathOperator{\divi}{div}

\begin{document}

\setpagewiselinenumbers
\title{A semi-analytical approach to cosmic void evolution}

\author[a]{A.N. Baushev}
\emailAdd{baushev@gmail.com}

\affiliation[a]{Bogoliubov Laboratory of Theoretical Physics, Joint Institute for Nuclear Research (JINR),\\
141980 Dubna, Moscow Region, Russia}

\author[b]{A.G. Nikiforov}
\emailAdd{anikiforov@inasan.ru}

\affiliation[b]{Institute of Astronomy, Russian Academy of Sciences,\\
119017 Moscow, Russia}

\author[b]{M.V. Barkov}
\emailAdd{barkov@inasan.ru}

\abstract{
We present a theoretical study of the non‑linear evolution of cosmic voids --- underdense regions that occupy a large fraction of the observable Universe. We model a void as an isolated homogeneous spheroidal (axisymmetric) ellipsoid embedded in a homogeneous $\Lambda$CDM universe. Starting from a small initial density contrast at redshift $z=500$, we numerically integrate the equations of motion for the ellipsoid semi‑axes and follow their evolution to the present epoch. We examine the anisotropic expansion of the void and the corresponding change in its shape, characterised by the eccentricity $e$. 
We find that the void non-sphericity always decreases, but rather slowly:  the eccentricity drops from $e\approx0.87$ at $z=500$ to $e\approx0.81$ at $z=0$. Thus the void becomes rounder but remains aspherical throughout the evolution. The evolution and final value of the void underdensity are virtually independent of the void's eccentricity. The nonlinearity of void evolution becomes apparent very early: a ten percent deviation from the linear regime occurs already at $z\simeq 8$, when $\varepsilon=\Delta\rho/\rho\sim 10\%$. Notably, our calculations show that the majority of voids are not strongly underdense and contain a significant amount of matter, $\mu>0.5$. 
}


\maketitle

\section{Introduction}
Cosmic voids---large underdense regions spanning tens to hundreds of megaparsecs---are the most prominent features of the large‑scale structure of the observable Universe \cite{bootesvoid, voidobservation2015}. They contain very few galaxies (and no giant galaxies or galaxy clusters), and their low matter density makes them sensitive probes of the background cosmology and of the physics of structure formation \cite{fang2019c, voidreview2026}. In the standard $\Lambda$CDM paradigm, voids form from small initial negative density perturbations that expand faster than the Hubble flow, gradually emptying out \cite{26, 27}. Since voids occupy a large fraction of the Universe's volume, their statistical properties and internal dynamics provide independent constraints on dark energy, modified gravity, and the nature of the initial fluctuations.

To do this, however, we need a reliable model of void formation for various sets of cosmological parameters. Unfortunately, the analytical approach is applicable mainly to the spherically symmetric case \cite{bertschinger1985, blumenthal1992, weygaert1993}. On the other hand, N-body simulations allow us to consider a wide variety of void geometries and to take tidal effects into account. However, simulations suffer from significant numerical effects \cite{vanderbosch2018, 21}. In a number of cases, obvious contradictions arise between simulation results and analytical estimates. For example, simulations predict that voids are practically empty: the matter density in them is $\sim 10$ times lower than the average Universe \cite{modelvoid, nbodyvoids2019}. In other words, their underdensity is $-\varepsilon \equiv (\rho_{M, void}-\bar\rho_M)/\bar\rho_M\simeq -0.9$; here we denote the relative underdensity by $\varepsilon$. However, analytical estimates \cite{26, 27} show that voids are far from empty, with $\varepsilon \simeq 0.5$. Observations do not allow us to measure $\varepsilon$ directly: we can observe only fairly large galaxies, while voids can also contain numerous dim dwarf galaxies, cold gas, and virtually any amount of dark matter. Therefore, semi-analytical methods of void modeling are of interest.

In this article, we follow the method suggested in \cite{silk1979}: the authors modeled the evolution of the Local Supercluster as an isolated homogeneous triaxial ellipsoid embedded in a homogeneous Universe. By analogy, the evolution of an isolated void can be modeled as a homogeneous triaxial ellipsoidal underdensity embedded in an expanding background.  As in \cite{silk1979}, we assume (although this is not entirely true) that the Universe outside the underdensity remains homogeneous and unperturbed; the boundary of the underdensity remains ellipsoidal, and the matter inside this ellipsoid is also uniformly distributed (see the rationale in \cite{silk1979}).

\begin{figure*}[ht]
  \subfigure[$\varepsilon_0 = 0.002$]{\includegraphics[scale=0.612]{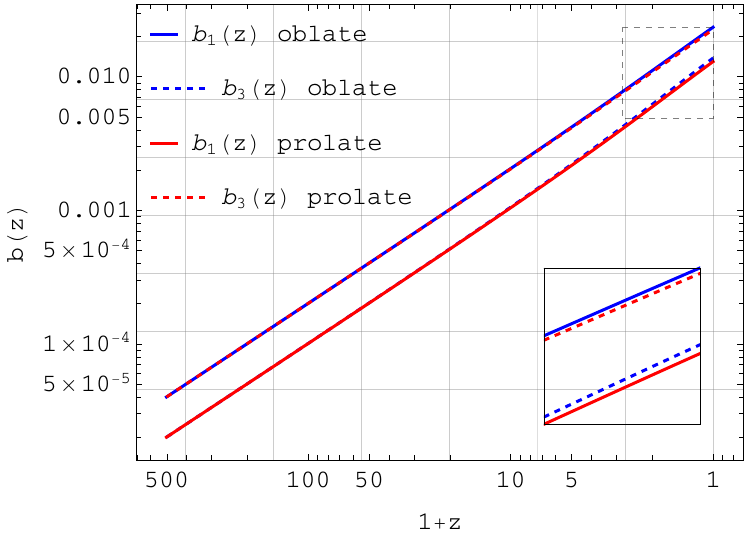}} \quad
  \subfigure[$\varepsilon_0 = 0.2$]{\includegraphics[scale=0.6]{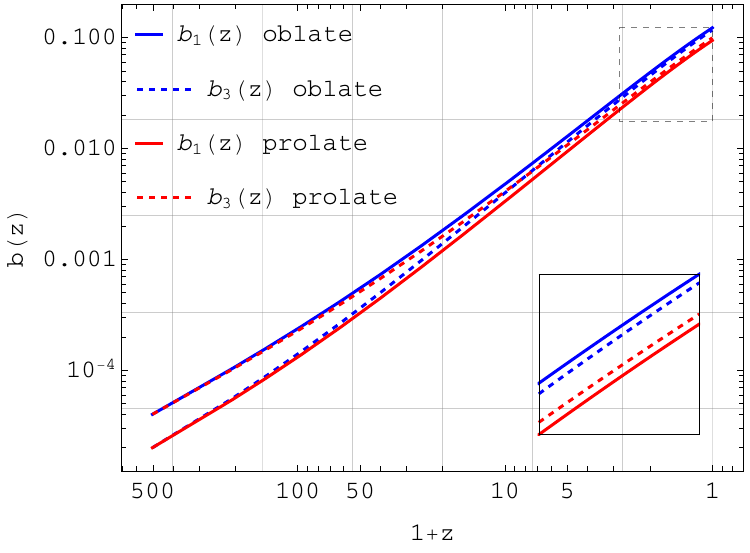}}
  \caption{Evolution of the equatorial semi‑axis $b_1$ and the symmetry semi‑axis $b_3$ of the underdense ellipsoidal region as functions of redshift $z$. Both axes grow roughly as $a(t)\propto(1+z)^{-1}$, but $b_1$ expands faster than $b_3$ because of the stronger outward gravitational pull in the equatorial plane of a prolate void. The logarithmic axes emphasise the Hubble‑like overall increase.}
  \label{fig:bdynamics}
\end{figure*}

The ellipsoidal model captures the essential non‑linear effects of self‑gravity, while remaining computationally tractable. In this work, we adopt such a model to follow the non‑linear evolution of a single underdense region from a high redshift, deep in the matter dominated era, until the present epoch. Although the general ellipsoidal model allows for three distinct semi‑axes $b_1\neq b_2\neq b_3$, in this work we restrict ourselves to the axisymmetric (spheroidal) case with $b_1=b_2\neq b_3$. This simplification is motivated by several reasons. First, a spheroidal void captures the qualitatively important features of anisotropic void expansion---namely, the difference between the growth of the equatorial and polar axes---while reducing the number of dynamical equations from three to two. Second, from a theoretical standpoint, the velocity field inside a proto‑void tends to stretch the region in a quasi‑planar manner, favouring a configuration with two equal axes \cite{27}. The resulting shape for an underdense region is either prolate, with the symmetry axis $b_3$ shorter than the equatorial axes $b_1$, or oblate, with the symmetry axis $b_3$ longer than the equatorial axes $b_1$. Finally, the spheroidal approximation has been widely used in analytic and numerical studies of void evolution, providing a convenient benchmark for comparing non‑linear effects with linear predictions.

One of our main goals is to quantify the departure of the void expansion from the linear perturbation theory prediction. To this end, we solve numerically the equations of motion for the semi‑axes of the ellipsoid, taking into account the exact gravitational potential coefficients and the time‑dependent background expansion. We set the initial conditions of the void at redshift $z=500$ and follow its evolution to $z=0$. The results are compared with the growth factor $D(a)$ of the linearized theory, allowing us to reveal the non‑linear effects of the underdensity evolution at late times. Throughout the paper, the subindices 'p' and '0' denote the values of quantities at the present time and at the moment $z=500$, respectively.

The paper is organized as follows. In Section~\ref{sec:model} we describe the ellipsoidal model and the equations of motion. Section~\ref{sec:results} presents the numerical results for the void shape, the eccentricity, and the density contrast, and discusses the deviation from linear theory. Our conclusions are summarized in the final section.


\begin{figure}
    \centering
    \includegraphics[width=0.99\linewidth]{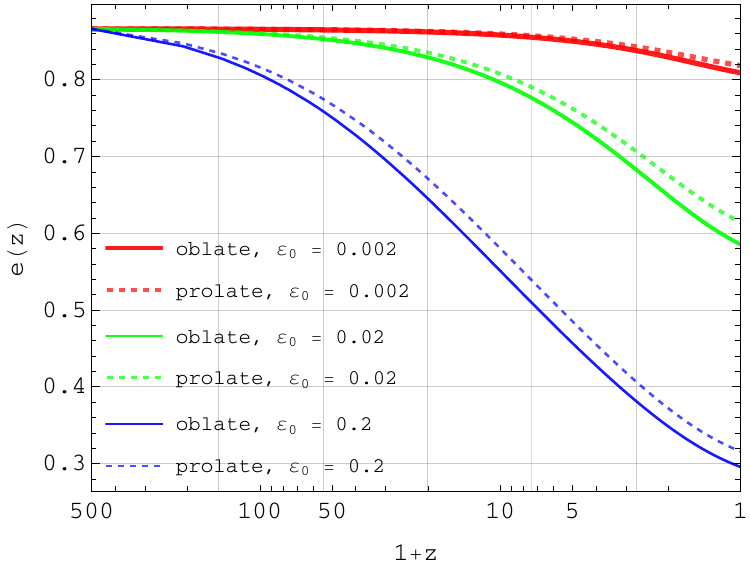}
    \caption{Void eccentricity $e(z)$ as a function of redshift. Starting from $e\approx0.866$ at $z=500$, the eccentricity decreases monotonically to $e\approx0.808$ at $z=0$ for $\varepsilon=0.002$. A higher initial density contrast value results in a faster decay of eccentricity. The void therefore becomes rounder with time, although it remains triaxial throughout the whole evolution.}
    \label{fig:eccentricity}
\end{figure}

\section{The evolution of the underdense ellipsoidal region}
\label{sec:model}
We consider the evolution of an isolated underdense region (void) in the standard $\Lambda$CDM Universe with present‑day matter density parameter $\Omega_{m,p}=0.306$ and dark energy density parameter $\Omega_{\Lambda,p}=1-\Omega_{m,p}$. We neglect the influence of the radiation component. Solving the Friedmann equation, we find that the background expansion is described by the scale factor
\begin{equation}
a(t)=a_p\left(\frac{\Omega_{m,p}}{\Omega_{\Lambda,p}}\right)^{1/3} 
\sinh^{2/3}\!\left(\frac{3}{2}\sqrt{\Omega_{\Lambda,p}}\,t\right),
\label{eq:scalefactor}
\end{equation}
where $t$ is the cosmic time in units of $H_p^{-1}$ ($a_p$ and $H_p$ are the present-day values of the scale factor and of the Hubble constant, respectively). Since the Universe is flat, we may set $a_p\equiv1$.  
The initial time $t_0$ corresponds to $z(t_0)=500$, which gives $a(t_0)\simeq 0.002$; the evolution is followed until the present epoch $z=0$. We denote the density contrast of the void by $\varepsilon(z)\equiv |\rho_{M, void}(z)-\bar\rho_M(z)|/\bar\rho_M(z)$, where $\rho_{M, void}$ and $\bar\rho_M$ are the matter densities inside the void and in the Universe on average, respectively, and set $\varepsilon_0\equiv\varepsilon(z=500)=0.002$.

The void is modeled as a homogeneous triaxial ellipsoid with semi‑axes $b_1(t)$ (the two equal axes in the plane of symmetry) and $b_3(t)$ (the symmetry axis).  
The evolution of the semi‑axes is governed by the equations of motion for an ellipsoid in an expanding background \cite{silk1979}, taking into account the self‑gravity of the ellipsoid, the tidal effects from the surrounding homogeneous matter distribution, and the cosmological constant:
\begin{align}
\ddot b_1 &= \left[\frac{\Omega_{m,p}}{a^3}
\left( \frac{3}{4}\varepsilon A(e) - \frac12 \right)
+ \Omega_{\Lambda,p} \right] b_1, \label{eq:b1ddot} \\
\ddot b_3 &= \left[\frac{\Omega_{m,p}}{a^3}
\left( \frac{3}{4} \varepsilon B(e) - \frac12 \right)
+ \Omega_{\Lambda,p} \right] b_3, \label{eq:b3ddot}
\end{align}
where the dots over variables denote the derivatives with respect to $t$.  
For a homogeneous ellipsoid of eccentricity $e=\sqrt{1-(b_3/b_1)^2}$, the gravitational potential coefficients $A(e)$ and $B(e)$ are derived in \cite{Lin1965} and corrected in \cite{28}:
\begin{align}
A(e)=\frac{\sqrt{1-e^2}}{e^3}\arcsin e - \frac{1-e^2}{e^2},\\
B(e)=\frac{2}{e^2} - \frac{2\sqrt{1-e^2}}{e^3}\arcsin e.
\end{align}
  
The initial eccentricity for a prolate spheroid (with axes $b_1 = b_2 < b_3$) is defined as $e=\sqrt{1-(b_1/b_3)^2}$. The shape coefficients $A(e)$ and $B(e)$ are given by
\begin{align}
 A(e) =\dfrac{1}{e^2}\left[1 - \dfrac{1-e^2}{2e}\ln\!\left(\dfrac{1+e}{1-e}\right)\right], \\
 B(e) =\dfrac{2(1-e^2)}{e^3}\left[\dfrac{1}{2}\ln\!\left(\dfrac{1+e}{1-e}\right) - e\right],
\end{align}
with the limiting values $A(0)=B(0)=2/3$ for both cases.

We denote by $\mu(t)$ the ratio of the matter density inside the void to the mean matter density of the Universe. It is related to $\varepsilon$ by a very simple relation:
\begin{equation}
\mu \equiv  (1-\varepsilon).
\label{eq:mu0}
\end{equation}
From mass conservation it follows that
\begin{equation}
\mu(t) = \left(\frac{a(t)}{a(t_0)}\right)^3 (1-\varepsilon_0)\,
\frac{b_{1}^2(t_0)\,b_{3}(t_0)}{b_{1}^2(t)\,b_{3}(t)} .
\label{eq:mu}
\end{equation}

The initial velocities inside the void are set consistently with the fact that the void is a linear growing mode in the matter-dominated Universe \cite{gorbrub2}:
\begin{align}
\dot b_1(t_0) &= b_{1}(t_0)\,
\sqrt{\frac{\Omega_{m,p}}{a^3(t_0)}+\Omega_{\Lambda,p}}\,
\left(1+\frac{1}{2}A(e_0)\,\varepsilon_0\right), \\
\dot b_3(t_0) &= b_{3}(t_0)\,
\sqrt{\frac{\Omega_{m,p}}{a^3(t_0)}+\Omega_{\Lambda,p}}\,
\left(1+\frac{1}{2}B(e_0)\,\varepsilon_0\right),
\end{align}
where the initial eccentricity is $e_0=\sqrt{1-\bigl(b_{3}(t_0)/b_{1}(t_0)\bigr)^2}$.

The system of equations (\ref{eq:b1ddot})--(\ref{eq:b3ddot}) is integrated numerically from $t_0$ to the cosmic time corresponding to $z=0$.

\section{Results and discussion}
\label{sec:results}
The numerical integration of the equations of motion (\ref{eq:b1ddot})--(\ref{eq:b3ddot}) yields the full non‑linear evolution of the void semi‑axes, the eccentricity, and the underdensity. Fig.~\ref{fig:bdynamics} shows the evolution of the semi‑axes $b_1$ and $b_3$.
We note that, since the volume scales as $V \propto b_1^2 b_3$ for both spheroidal cases, these particular ratios result in different initial volumes for the two morphologies. Both axes increase over time, following roughly the Hubble expansion, but the equatorial axis $b_1$ expands noticeably faster than the symmetry axis $b_3$. This differential growth is a result of the anisotropic gravitational forces inside the ellipsoid: matter is pulled outward more effectively along the plane of the two equal axes. The effect remains visible at all redshifts and leads to a systematic change in the void shape: it makes the void rounder. Thus, the void anisotropy decreases with time.

The evolution of the void shape is more directly illustrated by the eccentricity $e(z)$, plotted in Fig.~\ref{fig:eccentricity}. Initially $e=0.866$, and it drops steadily to $0.808$ at $z=0$. The decrease reflects the fact that $b_1$ grows faster than $b_3$, making the axis ratio $b_3/b_1$ smaller; thus $e$ also decreases. However, the change is modest: the void becomes only slightly rounder, suggesting that the initial triaxiality is largely preserved during the void's evolution. For a more extreme case with $\varepsilon_0=0.2$, the eccentricity drops significantly, reaching almost $e \approx 0.3$ by $z=0$, although such a large value of $\varepsilon$ exceeds the typical range expected from $\Lambda$CDM initial conditions; see Sec.~\ref{subsec:voidab} for details.

\begin{figure}[h]
\centering
\includegraphics[width=0.99\linewidth]{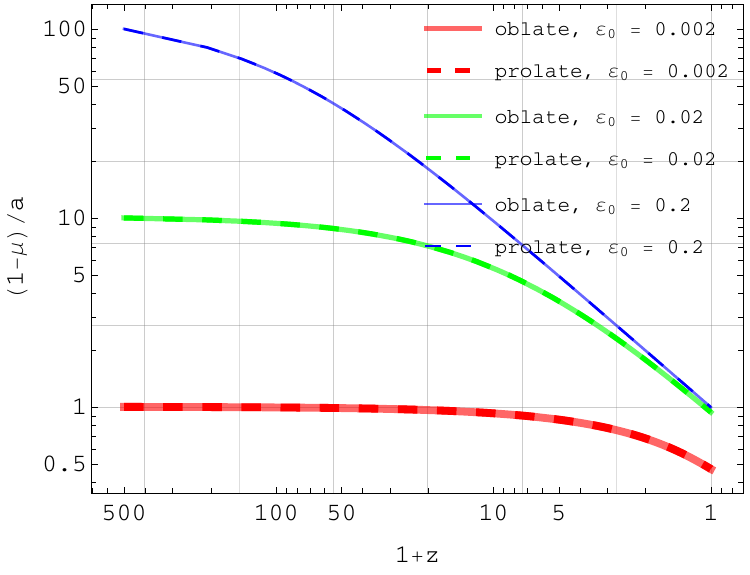}
\caption{Amplitude of the underdensity $\varepsilon/a\equiv(1-\mu)/a$ as a function of redshift. Initially set by $\varepsilon_0/a(t_0)\approx1$, the amplitude grows during the matter‑dominated era and reaches $\approx0.5$ at $z=0$. The growth is close to a power law but deviates at late times because of the influence of dark energy and the non‑linear coupling between the void expansion and the background cosmology.}
\label{fig:onemuscale}
\end{figure}

\begin{figure}[h]
\centering
\includegraphics[width=0.99\linewidth]{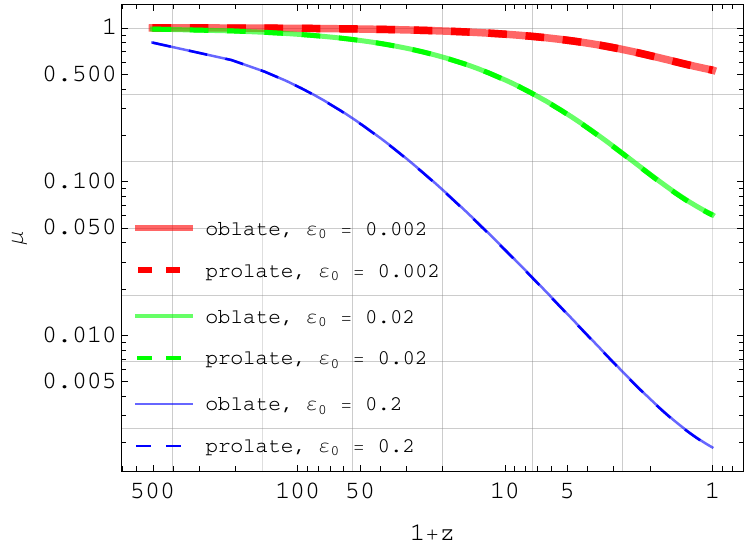}
\caption{Evolution of the relative density parameter $\mu$ as a function of redshift for spheroidal voids with oblate (solid lines) and prolate (dashed lines) geometries. 
Results are shown for three values of the initial density perturbation amplitude $\varepsilon_0 \in \{0.002,\,0.02,\,0.2\}$. 
All models start from $\mu(z=500) = 1-\varepsilon_0$.}
\label{fig:muoneplusz}
\end{figure}

A convenient measure of the underdensity amplitude is the quantity $\varepsilon/a$, plotted in Fig.~\ref{fig:onemuscale}. By construction, $\varepsilon/a=\varepsilon_0/a(t_0)\approx 1, 10, 100$ for $\varepsilon_0=0.002, 0.02, 0.2$, respectively, at the initial time, and it then decreases monotonically as the void empties out relative to the background (see Fig.~\ref{fig:muoneplusz}). At $z=0$ the amplitude reaches $\approx0.6$. The growth approximately follows the expected behavior for linear perturbations during the matter‑dominated epoch, but at $z\lesssim 10$ the slope gradually decreases, which, in principle, may be a consequence either of the non‑linear effects of void evolution or of the influence of dark energy. For larger initial perturbations, the deviation from the flat initial solution occurs at earlier times due to non‑linear effects.

Of course, the contribution of dark energy is still too small at $z=10$ to significantly influence void evolution. To show this, we introduce the linear growth factor $D(a)$ defined as
\begin{equation}
D(a) = \frac{\bigl(\Omega_{\Lambda,p} a^3 + \Omega_{m,p}\bigr)^{1/2}}{a^{3/2}}
\int_0^a \frac{b^{3/2}}{\bigl(\Omega_{\Lambda,p} b^3 + \Omega_{m,p}\bigr)^{3/2}}\,db .
\label{eq:Dgrowth}
\end{equation}
This factor describes the growth of matter perturbations in the $\Lambda$CDM model in the linear approximation \cite{loebreview}. Now we may compare the obtained density contrast with the predictions of linear theory. In linear theory, a small underdensity would evolve as $\varepsilon(t)\propto D(a)$, so that $\varepsilon/a$ should track $D(a)/a$ if the evolution remains linear.

\begin{figure}[h]
\centering
\includegraphics[width=0.99\linewidth]{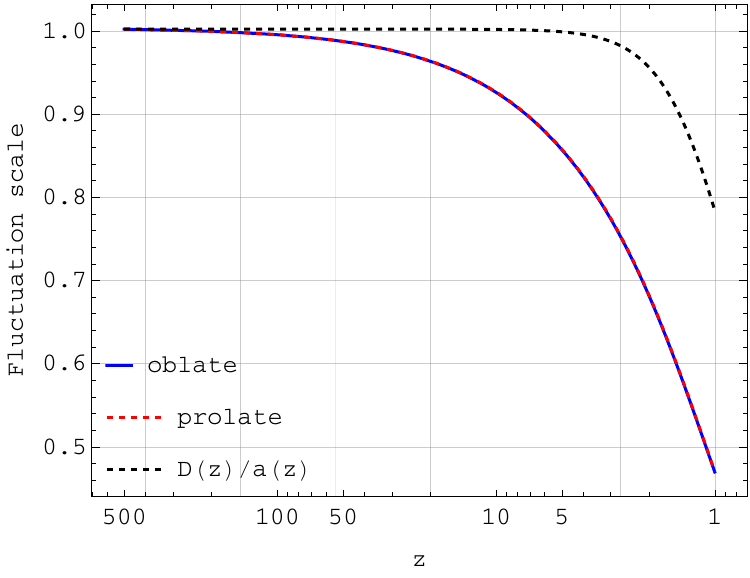}
\caption{Comparison of the non‑linear underdensity $\varepsilon/a$ for oblate and prolate cases with the linear growth factor $D(a)/a$ (dashed black line) as functions of cosmic time $\tau=H_p t$ for initial $\varepsilon_0=0.002$. At early times ($\tau\lesssim 0.1$, $z\gtrsim10$) the two curves coincide, confirming the validity of the linear approximation. At later times the non‑linear amplitude falls behind the linear extrapolation; by today the ratio is about $0.6$.}
\label{fig:muAndD}
\end{figure}

Fig.~\ref{fig:muAndD} shows both $(1-\mu)/a$ and $D(a)/a$ as functions of the cosmic time $\tau=H_p t$. The two curves practically overlap at $\tau\lesssim 0.1$ ($z\gtrsim 10$), confirming that the initial evolution is well described by linear theory. At later times, however, the non‑linear amplitude systematically falls behind the linear extrapolation. At the present epoch, $(1-\mu)/a$ is only about $60\,\%$ of the value that linear evolution would predict. 

The departure from linear growth is summarized by the ratio $\xi(z)\equiv D/(1-\mu)$, displayed in Fig.~\ref{fig:muToD}. By definition, $\xi=1$ for a strictly linear perturbation, and it increases as the non‑linear evolution slows down relative to the linear prediction. Starting from unity at $z=500$, $\xi$ grows steadily and reaches $\approx1.7, 8.4, 80$ by $z=0$ for $\varepsilon_0=0.002, 0.02, 0.2$, respectively. Thus, the actual underdensity inside the void is about $40\,\%$ smaller than what would be inferred from extrapolating linear theory. Such a non‑linear correction is essential for properly interpreting observations of large voids in galaxy surveys and for comparing them with theoretical predictions of the $\Lambda$CDM model.

\subsection{The probability of void formation}

The non‑linear relationship between the initial density contrast $\varepsilon_0$ and the present‑day void depth $\mu=1-\varepsilon\equiv\rho_v/\bar\rho$ is the key ingredient for converting the Gaussian statistics of primordial fluctuations into a prediction for the abundance of cosmic voids. As we have seen, the ellipsoidal modeling of voids requires solving a system of two coupled differential equations for each value of $\varepsilon$, which is computationally expensive when a large number of values is needed. It is therefore natural to ask whether a simpler spherical void model can reproduce the ellipsoidal results with sufficient accuracy.

For a perfectly spherical uniform underdense region, the dynamics reduces to the evolution of the dimensionless radius $p(\tau)=R(\tau)/R(\tau_0)$, where $R(\tau)$ is the physical radius of the sphere. Because of the spherical symmetry of the system and the spatial constancy of the density inside the void, the underdense region may be described by the Friedmann solution. The equation of motion, analogous to the Friedmann equation for an open universe with a negative curvature term, reads:
\begin{equation}
\frac{dp}{d\tau}=p\,\sqrt{\,\Omega_{m,p}(1+z_0)^3
\left(\frac{1-\varepsilon_0}{p^3}+\frac{5}{3}\frac{\varepsilon_0}{p^2}\right)+\Omega_{\Lambda,p}}\,,
\label{eq:sphere}
\end{equation}
with the initial condition $p(\tau_0)=1$ (for the derivation of equation~\ref{eq:sphere}, see Appendix~\ref{appen}). This equation can be integrated directly. The density ratio at any time is then given by mass conservation:
\begin{equation}
\mu(\tau)=(1-\varepsilon_0)\left(\frac{a(\tau)}{a(\tau_0)}\right)^3\frac{1}{p^3(\tau)} .
\label{eq:mu_sphere}
\end{equation}
In the limit $\varepsilon_0\to0$, Eq.~\eqref{eq:sphere} describes the linear growth of a small spherical underdensity and is exactly equivalent to the linear perturbation theory result.

\begin{figure}[t]
\centering
\includegraphics[width=0.99\linewidth]{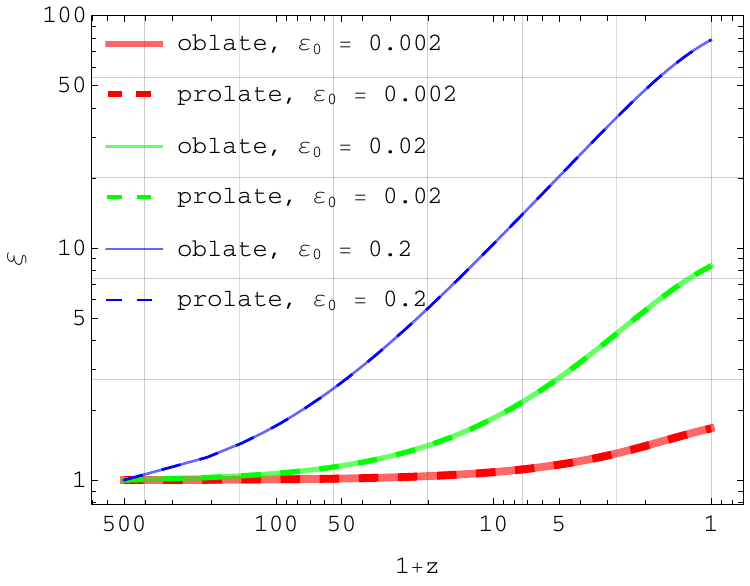}
\caption{Ratio $\xi(z)\equiv D(a)/\varepsilon$ quantifying the deviation from linear evolution. The ratio is unity at early times and increases steadily to $\xi\approx1.7, 8.4, 79.4$ for $\varepsilon_0 = 0.002,0.02,0.2$ respectively at $z=0$.}
\label{fig:muToD}
\end{figure}

\begin{figure}[t]
\includegraphics[width=0.99\linewidth]{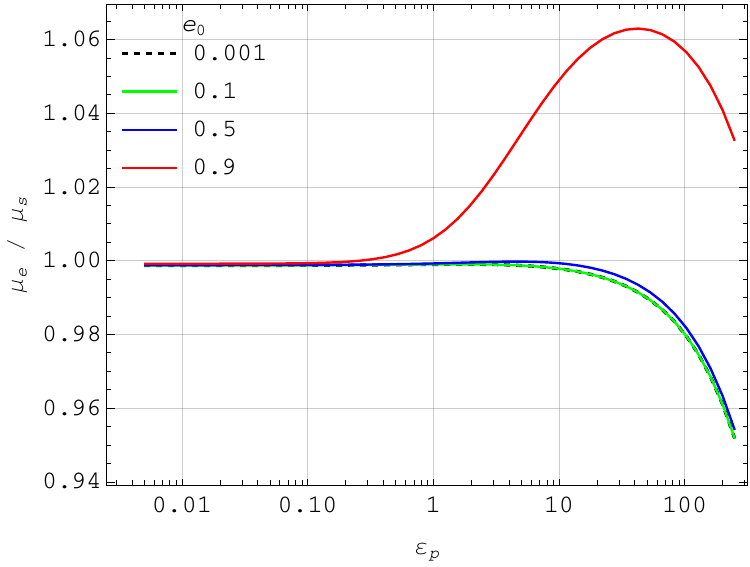}
\caption{Present‑day density contrast ratio $\mu_e/\mu_s$ as a function of the linearly extrapolated initial contrast $\varepsilon_p \equiv \varepsilon_0(1+z_0)$ for the spherical model ($e_0 = 0$) and for ellipsoidal models with several initial eccentricities $e_0$. All curves practically coincide, demonstrating that the spherical approximation accurately captures the non‑linear void depth for the range of $\varepsilon_0$ relevant to observed voids.}
\label{fig:mueps}
\end{figure}

Fig.~\ref{fig:mueps} shows the relation $\mu(\varepsilon_0)$ at $z=0$, obtained from the spherical model and from the ellipsoidal model for four different initial eccentricities $e_0=0.001,\,0.1,\,0.5,\,0.9$.
The initial contrast $\varepsilon_0$ at $z=500$ has been linearly rescaled to $z=0$ via $\varepsilon_p\equiv\varepsilon\,(1+z_0)$ to highlight the non‑linear suppression.
All curves lie almost exactly on top of each other, indicating that the shape of the void (its eccentricity) has a negligible influence on the final density contrast. This result is in perfect agreement with results presented erlier on Figs.~\ref{fig:onemuscale}--\ref{fig:muToD} for eccentrisity $\epsilon=0.866$.
These results justifies using the computationally much faster spherical model to calculate $\mu(\varepsilon_0)$ for statistical purposes.

\subsubsection{Void abundance}
\label{subsec:voidab}

According to the standard picture, the entire large-scale structure of the Universe formed from primordial Gaussian perturbations. The fraction of the Universe's volume occupied by fluctuations whose linearly extrapolated overdensity (or underdensity) exceeds a chosen threshold is then obtained from the statistics of the Gaussian initial density field \cite{loebreview, 1974ApJ...187..425P}. We consider the linear density contrast smoothed on a comoving scale $R$ and linearly extrapolated to $z=0$:
\begin{equation}
\delta_{R}(\mathbf{x})=\int \delta_{0}(\mathbf{x}')\,
W(|\mathbf{x}-\mathbf{x}'|;R)\,d^{3}x',
\label{eq:smoothed_delta}
\end{equation}
where $\delta_{0}(\mathbf{x})\equiv\delta(\mathbf{x},z)/D_{+}(z)$ is the density contrast evolved to the present epoch, $D_{+}(z)$ is the linear growth factor (normalized to unity at $z=0$), and $W(r;R)$ is a window function, commonly taken to be a top‑hat of radius $R$. The root mean square fluctuation of the smoothed field is
\begin{equation}
\sigma_{R}^{2}\equiv\langle\delta_{R}^{2}\rangle
=\int\frac{d^{3}k}{(2\pi)^{3}}\,P(k)\,|\tilde{W}(kR)|^{2},
\label{eq:sigmaR}
\end{equation}
with $P(k)$ being the linear matter power spectrum at $z=0$ and
$\tilde{W}(kR)$ the Fourier transform of the window function.
For a top‑hat filter $\tilde{W}(x)=3(\sin x-x\cos x)/x^{3}$.

The probability that a randomly chosen point lies in a region whose linearly extrapolated contrast exceeds $\varepsilon_{0}$ is
\begin{equation}
    \begin{split}
            \mathcal{P}(>\varepsilon_{0})=\frac{1}{\sqrt{2\pi}}
            \int_{\varepsilon_{0}}^{\infty} \frac{d\delta_{R}}{\sigma_{R}}
            \exp\!\left(-\frac{\delta_{R}^{2}}{2\sigma_{R}^{2}}\right)=\\
            =\frac12\,\mathrm{erfc}\!\left(\frac{\varepsilon_{0}}{\sqrt{2}\sigma_{R}}\right).
            \label{eq:prob_ps}
    \end{split}
\end{equation}

\begin{figure}[t]
\centering
\includegraphics[width=0.99\linewidth]{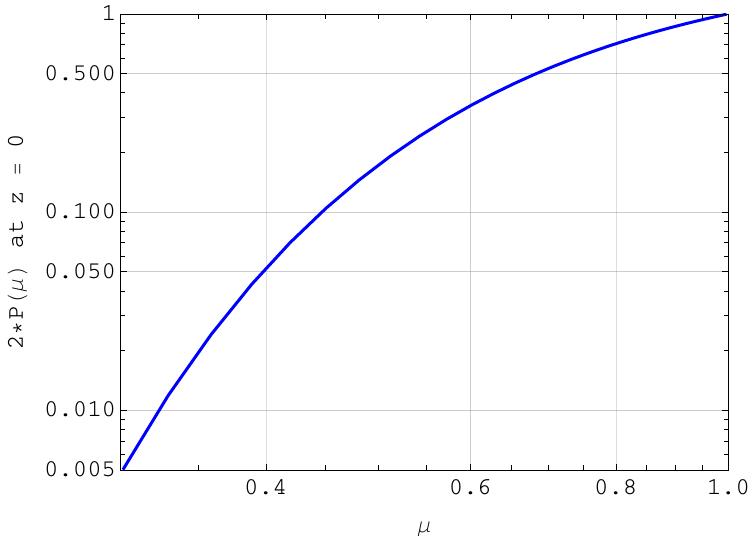}
\caption{Cumulative volume fraction of voids with present‑day density contrast smaller than $\mu$ (i.e. deeper underdensities), computed using the spherical model mapping. The curve is shown for a smoothing scale $R=8\,h^{-1}\,\mathrm{Mpc}$ and the standard $\Lambda$CDM power spectrum.}
\label{fig:Ptomu}
\end{figure}

Equation~\eqref{eq:prob_ps} gives the volume fraction of the Universe that lies above the threshold (or below it for underdensities, since the Gaussian distribution is symmetric).  The additional factor of $2$ that appears in the final expression $2\mathcal{P}(\mu)$ is the ``fudge factor'' that compensates for the cloud‑in‑cloud problem and ensures that the entire volume is assigned to either voids or collapsed objects. Hence, the cumulative volume fraction occupied by voids whose present‑day density contrast is smaller than $\mu$ (i.e., underdensities deeper than $\mu$) is
\begin{equation}
2\mathcal{P}(\mu)=2\,\mathcal{P}\!\big(\varepsilon_{0}(\mu)\big)
=\mathrm{erfc}\!\left(\frac{\varepsilon_{0}(\mu)}{\sqrt{2}\,\sigma_{R}}\right),
\label{eq:Pmu_final}
\end{equation}
where $\varepsilon_{0}(\mu)$ is the inverse of the non‑linear mapping $\mu(\varepsilon_{0})$ obtained from the spherical (or ellipsoidal)
model.

In the above, $\sigma_{R}$ is evaluated at $z=0$; the smoothing radius $R$ is fixed by the physical scale of the voids one intends to describe.  For the analysis presented in this paper we adopt $R=8\,h^{-1}\,\mathrm{Mpc}$, a scale typical for moderate voids detected in galaxy redshift surveys.  The corresponding $2\mathcal{P}(\mu)$
is displayed in Fig.~\ref{fig:Ptomu} and shows that voids with $\mu \lesssim 0.6$ occupy only a quarter of the cosmic volume. 

Observations of voids in large galaxy redshift surveys indicate that most voids are extremely empty: as shown by \cite{Hoyle2004} for the 2dFGRS, the mean density contrast of voids is
$\Delta_m\simeq-94\%$, corresponding to a residual matter density $\mu \simeq0.06$.  A more moderate threshold $\mu \simeq 0.2$ ($\Delta_m\simeq-80\%$) has been adopted in theoretical works based on the excursion‑set formalism (e.g., \cite{Jennings2013}). Our results suggest a less pronounced underdensity, $\mu \sim 0.5$, and this result is supported by analytical models of the central void density~\cite{26, 27}.  

\section{Conclusions}
We conducted a study of the evolution of an individual void using a semi-analytical method. The main results can be summarized as follows:

\noindent 1) As expected from general theoretical considerations, the nonsphericity of the underdensity decreases during its evolution. This effect is of the same nature as the increase in nonsphericity during the evolution of an overdensity. However, an important difference is found: in the overdensity case, the nonsphericity grows rapidly, and the ellipsoid transforms into a pancake. By contrast, in the case of a void, the reduction in non-sphericity proceeds very slowly.

\noindent 2) The evolution and final value of the void underdensity are virtually independent of the void's eccentricity. The theoretical reasons for this are given in~\cite{27}.

\noindent 3) The nonlinearity of void evolution becomes apparent very early. Figure~\ref{fig:muToD} shows that a ten percent deviation from the linear regime occurs already at $z\simeq 8$, when $\varepsilon=\Delta\rho/\rho\sim 10\%$.

\noindent 4) Our semi-analytical calculation confirms that voids are not so empty, $\Delta_m \simeq -50\%$ ($\mu>0.5$). This result is in agreement with theoretical estimates~\cite{26, 27}, but contradicts the results of N-body simulations.

\bibliographystyle{JHEP}
\bibliography{article_biblio}

\appendix
\section{Derivation of equation~(\protect{\ref{eq:sphere}})}
\label{appen}

We use the method proposed in~\cite{24} to derive equation~(\ref{eq:sphere}). Let us consider a homogeneous flat Friedmann universe with $\Omega_{m,p}=0.306$, $\Omega_{\Lambda}+ \Omega_{m,p}=1$. There is a single small uniform spherically symmetric underdensity with initial density contrast $\varepsilon_0$ at redshift $z_0\gg 10$. We set the origin of coordinates at the center of the underdensity. 

Here we denote the Universe density at $z=z_0$ by $q$. The density inside the underdensity does not depend on the coordinates and is equal to $q-\Delta q= q- \varepsilon_0 q$ at $z=z_0$. Because of the spherical symmetry of the system, the underdense region may be described by the Friedmann metric, but with different parameters (a detailed justification is given in~\cite{24}). The Hubble constant of the underdensity is equal to $H+h$; we denote the matter, curvature, and dark energy densities by $\sigma_{m}, \sigma_{a}$, and $\sigma_\Lambda=\rho_\Lambda$, respectively. Our aim is to find the evolution of the dimensionless radius $p(t)=R(t)/R(t_0)\equiv R/R_0$, where $R(t)$ is the physical radius of the underdensity. It is fully determined by the Friedmann equation, but we need to determine
$\sigma_{m,0}, \sigma_{a,0}$, i.e., $\sigma_{m}, \sigma_{a}$ at $z_0$.

Let us write the continuity equation $\dfrac{d\rho}{dt}+\divi(\rho\vec v)=0$. For a Friedmann universe, $\divi(\rho\vec v)=3\rho H$. Then the continuity equations for the undisturbed Universe and for the underdense region have the form
\begin{eqnarray}
    &\dfrac{d\rho}{dt} = -3\rho H \nonumber\\
    &\dfrac{d\rho}{dt}-\dfrac{d\Delta\rho}{dt} =-3(H+h)(\rho-\Delta\rho)
\end{eqnarray}
Combining these equations and neglecting second-order terms, we obtain:
\begin{equation}
\label{28app2}
3h\rho=3H\Delta\rho+\frac{d\Delta\rho}{dt}.
\end{equation}
We consider our system at $z_0$, deep in the matter-dominated stage, and therefore $\Delta\rho/\rho\propto a$. Differentiating this equation, we obtain
\begin{equation}
\label{28app3}
d\Delta\rho=\Delta\rho\left(\dfrac{d\rho}{\rho}+\dfrac{da}{a}\right)
\end{equation}
Since $\rho\propto a^{-3}$ in the matter-dominated stage, $\dfrac{d\rho}{\rho}=-3\dfrac{da}{a}=-3Hdt$ and
\begin{equation}
\label{28app4}
\dfrac{d\Delta\rho}{dt}=-2\Delta\rho H.
\end{equation}
Substituting this result into~(\ref{28app2}), we obtain
\begin{equation}
\label{28app5}
h=\dfrac{\Delta\rho}{3\rho}H=\dfrac{\varepsilon}{3}H.
\end{equation}
The general form of the Friedmann equation for the underdense region is
\begin{equation}
\label{28app6}
\left(\dfrac{dR}{Rdt}\right)^2=\dfrac{8\pi G}{3}\left[\sigma_{m,0}\left(\frac{R_0}{R}\right)^3+\sigma_{a,0}\left(\frac{R_0}{R}\right)^2+\sigma_\Lambda\right]
\end{equation}
Substituting here $p(t)= R/R_0$ and taking into account that $\sigma_\Lambda=\rho_\Lambda={\it const}$, we obtain
\begin{equation}
\label{28app7}
\left(\dfrac{dp}{pdt}\right)^2=\dfrac{8\pi G}{3}\left[\dfrac{\sigma_{m,0}}{p^3}+\dfrac{\sigma_{a,0}}{p^2}+\rho_\Lambda\right]
\end{equation}
where
\begin{equation}
\label{28app8}
\dfrac{dR}{Rdt}=\dfrac{dp}{pdt}=H+h
\end{equation}
is the Hubble constant inside the underdensity. Let us rewrite~(\ref{28app7}) at $z=z_0$, where
\begin{eqnarray}
\label{28app9}
 \sigma_{m,0}= \rho_{m,0} (1-\varepsilon_0), \qquad \rho_\Lambda\ll\rho_{m,0},\\
 (H+h)^2=H^2\left(1+\dfrac{\varepsilon_0}{3}\right)^2\simeq\dfrac{8\pi G}{3}\rho_{m,0} \left(1+\frac23\varepsilon_0\right)
 \label{28app10}
\end{eqnarray}
We used equation~(\ref{28app5}) to derive~(\ref{28app10}). Substituting~(\ref{28app9}-\ref{28app10}) into~(\ref{28app7}) and simplifying, we get
\begin{equation}
\label{28app11}
\rho_{m,0} \left(1+\frac23\varepsilon_0\right)=\rho_{m,0}(1-\varepsilon_0)+\sigma_{a,0}.
\end{equation}
Thus
\begin{equation}
\label{28app12}
\sigma_{a,0}=\frac53 \varepsilon_0\rho_{m,0}.
\end{equation}
Now we substitute this value and the value~(\ref{28app9}) for $\sigma_{m,0}$ into~(\ref{28app7}):
\begin{equation}
\label{28app13}
\left(\dfrac{dp}{pdt}\right)^2=\dfrac{8\pi G}{3}\left[\dfrac{\rho_{m,0} (1-\varepsilon_0)}{p^3}+\frac53\dfrac{ \varepsilon_0\rho_{m,0}}{p^2}+\rho_\Lambda\right].
\end{equation}
Let us take into account that
\begin{eqnarray}
\label{28app14}
\dfrac{8\pi G}{3}\rho_{m,0}=\dfrac{\Omega_{m,p}}{(z_0+1)^3}H_p^2\\
\dfrac{8\pi G}{3}\rho_\Lambda=\Omega_{\Lambda,p} H_p^2 \label{28app15}
\end{eqnarray}
Substituting these values into~(\ref{28app13}), we finally obtain
\begin{equation}
\label{28app16}
\left(\dfrac{dp}{pdt}\right)^2=H_p^2\left[\Omega_{m,p}(1+z_0)^3
\left(\frac{1-\varepsilon_0}{p^3}+\frac53\frac{\varepsilon_0}{p^2}\right)+\Omega_{\Lambda,p}\right].
\end{equation}
One may see that this equation is equivalent to~(\ref{eq:sphere}).


\end{document}